\documentclass[11pt,leqno]{article}
\usepackage{cmu-titlepage2}
\usepackage{fullpage}
\usepackage[small,bf]{caption2}
\usepackage{times}
\usepackage{helvet}
\usepackage{courier}

\usepackage{subfigure}
\usepackage{cite} 
\usepackage{url,times}
\usepackage{epsfig,url}
\usepackage{subfigure}

\usepackage{amssymb}
\usepackage{amsmath}
\usepackage{amsfonts}

\newcommand{\ignore}[1]{}

\newtheorem{theorem}{Theorem}

\newtheorem{fact}{Fact}[section]
\newtheorem{definition}[fact]{Definition}

\newenvironment{proof}{\noindent{\bf Proof}:}{$\hfill
     \Box$\vspace*{0.02in}}

 \title{\hspace*{15mm}Shuffling a Stacked Deck: The Case for Partially \hspace*{15mm}Randomized Ranking of Search Engine Results \vspace*{10mm}}
 \author{\begin{tabular}{c c c}
Sandeep Pandey & Sourashis Roy & Christopher Olston\\
Carnegie Mellon University & UCLA & Carnegie Mellon University \\
\end{tabular} \\
\\
\begin{tabular}{c c}
Junghoo Cho & Soumen Chakrabarti\thanks{\hspace{2mm}This work was
performed while the author was visiting Carnegie Mellon University.}\\
UCLA & IIT Bombay
\end{tabular}
\vspace*{5mm}
}
 \date{March 2005}
 \trnumber{CMU-CS-05-116}
 \abstract{In-degree, PageRank, number of visits and other measures of Web page
popularity significantly influence the ranking of search results by
modern search engines.
The assumption is that {\em popularity} is closely correlated
with {\em quality}, a more elusive concept that is difficult to measure directly.
Unfortunately, the correlation between popularity and quality is very weak
for newly-created pages that have yet to receive many visits and/or
in-links.  Worse, since discovery of new content is largely done by
querying search engines, and because users usually focus their attention on the
top few results, newly-created but high-quality pages are effectively
``shut out,'' and it can take a very long time before they become
popular.  \\ \hspace*{5mm} We propose a simple and elegant solution to this problem: the
introduction of a controlled amount of randomness into search result
ranking methods. Doing so offers new pages a
chance to prove their worth, although clearly using too much
randomness will degrade result quality and annul any benefits
achieved.  Hence there is a tradeoff between {\em exploration} to estimate the
quality of new pages and {\em exploitation} of pages already known to be of
high quality.  We study this tradeoff both analytically and via simulation,
in the context of an economic objective function based on aggregate result
quality amortized over time.
We show that a modest amount of randomness leads to improved search results.
}

\begin{document}

\maketitle

\section{Introduction\label{sec:intro}}
Search engines are becoming the predominant means of discovering and accessing content on the Web.  
Users access Web content via a combination of following hyperlinks (browsing)
and typing keyword queries into search engines (searching).  Yet as
the Web overwhelms us with its size, users naturally turn to increased
searching and reduced depth of browsing, in relative terms.
In absolute terms, an estimated $625$ 
million search queries are received by major 
search engines each day~\cite{searchEngineWatch}.  

Ideally, search engines should
present query result pages in order of some intrinsic
measure of \emph{quality}. Quality cannot be measured directly.
However, various notions of \emph{popularity}, such as number of
in-links, PageRank~\cite{pagerank}, number of visits, etc., can be measured.  Most Web
search engines assume that popularity is closely correlated with
quality, and rank results according to popularity.

\subsection{The Entrenchment Problem}
Unfortunately, the correlation between popularity
and quality is very weak for newly-created pages that have few visits and/or
in-links.  Worse, the process by which new, high-quality pages accumulate popularity is
actually inhibited by search engines.
Since search engines dole out a limited number
of clicks per unit time among a large number of pages, always listing highly popular
pages at the top, and because users usually focus their attention on
the top few results~\cite{joachims,lempel}, 
newly-created but high-quality pages are ``shut out.''
This increasing ``entrenchment effect'' has witnessed broad commentary
across political scientists, the popular press, and Web researchers
\cite{sherman,power,bias,contro,googlearchy,define} and even led to the term
\emph{Googlearchy}.   In a recent study, Cho and Roy~\cite{impact} 
show that heavy reliance on a search engine that ranks results according
to popularity can delay widespread
awareness of a high-quality page by a factor of over $60$, compared with a
simulated world without a search engine in which pages are accessed through 
browsing alone.

Even if we ignore the (contentious) issue of fairness, there are
well-motivated economic objectives that are penalized by the entrenchment
effect.  Assuming a notion of intrinsic page quality as perceived by users,
a hypothetical ideal search engine would bias users
toward visiting those pages of the highest quality at a given time, 
regardless of popularity.  Relying on
popularity as a surrogate for quality sets up a vicious cycle of neglect
for new pages, even as entrenched
pages collect an increasing fraction of user clicks.  Given that some of
these new pages will generally have higher quality than some
entrenched pages, pure popularity-based ranking clearly fails to maximize
an objective based on average quality of search results seen by users.

\subsection{Entrenchment Problem in Other Contexts}

The entrenchment problem may not be unique to the Web search engine
context.  For example, consider recommendation
systems~\cite{kum-recom}, which are widely used in
e-commerce~\cite{berk-cob}.  Many users decide which items to view
based on recommendations, but these systems make recommendations based
on user evaluations of items they view.  This circularity leads to the
well-known \emph{cold-start} problem, and is also likely to lead to
entrenchment.

Indeed, Web search engines can be thought of as recommendation systems
that recommend Web pages.  The entrenchment problem is particularly
acute in the case of Web search, because the sheer size of the Web
forces large numbers of users to locate new content using search
engines alone.  Therefore, in this paper, we specifically focus on
diminishing the entrenchment bias in the Web search context.

\begin{figure}[!t]
\begin{center}
\includegraphics[width=180pt]{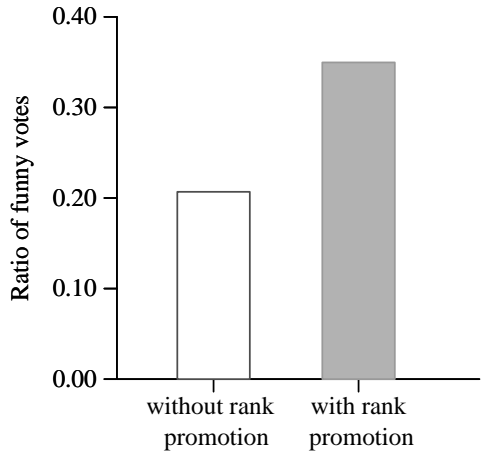}
\caption{Improvement in overall quality due to rank promotion in live study.}
\label{fig:intbarQPC}
\end{center}
\end{figure}

\subsection{Our Key Idea: Rank Promotion \label{sec:intro:jokes} }

We propose a very simple
modification to the method of ranking search results according to popularity:
promote a small fraction of unexplored pages up in the result list.
A new page now has some chance of attracting
clicks and attention even if the initial popularity of the page
is very small.  If a page has high quality, the rank boost gives the
page a chance to prove itself. 
(Detailed definitions and algorithms are given later in the paper.)

As an initial test for effectiveness, we conducted a 
real-world study, which we now describe briefly
(a complete description is provided in 
Appendix~\ref{sec:jokes}).
We created our own small Web community consisting of several thousand
Web pages, each containing a joke/quotation gathered from online databases.
We decided to use ``funniness'' as a surrogate
for quality, since users are generally willing to provide their opinion 
about how funny something is. Users had the option
to rate the funniness of the jokes/quotations they visit.
The main page of the Web site we set up consisted of an ordered list of links 
to individual joke/quotation pages, in groups of ten at a time, as is typical in 
search engine responses. Text at the top stated that the
jokes and quotations were presented
in descending order of funniness, as rated by users of the site.

A total of
$962$ volunteers participated
in our study over a period of $45$ days. Users were split at random into two user groups:
one group for which a simple form of rank promotion was used,
and one for which rank promotion was not used.
The method of rank promotion we used in this experiment is to place
new pages immediately below rank position 20.
For each user group we measured the ratio of funny votes to total votes during this period.
Figure~\ref{fig:intbarQPC} shows the result.  The ratio achieved using rank promotion
was approximately $60\%$ larger than that obtained using
strict ranking by popularity. 

\subsection{Design of Effective Rank Promotion Schemes}
\label{sec:introDesign}

In the search engine context
it is probably not appropriate to insert promoted pages
at a consistent rank position (lest users learn over time to avoid them).
Hence, we propose a simple \emph{randomized rank promotion} scheme in which promoted pages are assigned
randomly-chosen rank positions.

Still, the question remains as to how aggressively one should promote
new pages. Many new pages on the Web are not of high quality. Therefore, the extent of
rank promotion has to be limited very carefully, lest we negate the benefits of 
popularity-based ranking
by displacing pages known to be of high quality too often. 
With rank promotion there is an inherent tradeoff between 
{\em exploration} of new pages and {\em exploitation}
of pages already known to be of high quality.
We study how to balance these two aspects, in the context of
an overarching objective of maximizing the average quality of search
results viewed by users, amortized over time. 
In particular we seek to answer the following questions:

\begin{itemize}
\item Which pages should be treated as candidates for exploration, i.e., included 
in the rank promotion process so as to receive transient rank boosts?
\item Which pages, if any, should be exploited unconditionally, i.e., protected from 
any rank demotion caused by promotion of other pages?
\item What should be the overall ratio of exploration to exploitation?
\end{itemize}

Before we can begin to address these questions, we must model the
relationship between user queries and search engine results.
We categorize the pages on
the Web into disjoint groups by {\em topic}, such that each page pertains
to exactly one topic. Let $\mathcal{P}$ be the set of pages devoted to a particular 
topic $T$ (e.g., ``swimming'' or ``Linux''),
and let $\mathcal{U}$ denote the set of users interested in topic $T$.
We say that the users  $\mathcal{U}$ and pages $\mathcal{P}$
corresponding to topic $T$, taken together make up a {\em Web community}.
(Users may participate in multiple communities.)
For now we assume all users access the Web uniquely through a (single) search engine.
(We relax this assumption later in Section~\ref{sec:mixedbrowsing}.)
We further assume a one-to-one correspondence between queries and
topics, so that each query returns exactly the set of pages for the corresponding community.
Although far from perfect, we believe this model 
preserves the essence of the dynamic process we seek to understand.

Communities are likely to differ a great deal in terms of factors like
the number of users, the number of pages,
the rate at which users visit pages,
page lifetimes, etc.
These factors play a significant role in
determining how a given rank promotion scheme
influences page popularity evolution.
For example, communities with very active users
are likely to be less susceptible to the entrenchment effect
than those whose users do not visit very many pages.
Consequently, a given rank promotion scheme is bound to
create quite different outcomes in the two types of communities.
In this paper we provide an analytical method for predicting
the effect of deploying a particular randomized rank promotion
scheme in a given community, as a function of the most 
important high-level community characteristics.

\subsection{Experimental Study}

We seek to model a very complex dynamical system involving search
engines, evolving pages, and user actions, and trace its trajectory in
time.  It is worth emphasizing that even if we owned the most popular
search engine in the world, ``clean-room'' experiments would be
impossible.  We could not even study the effect of different choices
of a parameter, because an earlier choice would leave large-scale and
indelible artifacts on the Web graph, visit rates, and popularity of
certain pages.  Therefore, analysis and simulations are inescapable,
and practical experiments (as in Section~\ref{sec:intro:jokes}) must
be conducted in a sandbox.

Through a combination of analysis and simulation, we arrive at a
particular recipe for randomized rank promotion that 
balances exploration and exploitation effectively, and yields good
results across a broad range of community types.  Robustness is desirable
because, in practice, communities are not disjoint and therefore their
characteristics cannot be measured reliably.

\subsection{Outline}

In Section~\ref{sec:definitions} we present our model of Web page popularity,
describe the exploration/exploitation tradeoff as it exists in our context, 
and introduce two metrics for evaluating rank promotion schemes.
We then propose a randomized method of rank promotion in Section~\ref{sec:randomized},
and supply an analytical model of page popularity evolution under 
randomized rank promotion in Section~\ref{sec:TBP}.
In Sections~\ref{sec:experiments}--\ref{sec:mixedbrowsing}
we present extensive analytical and simulation results, 
and recommend and evaluate a robust recipe for randomized rank promotion.

\section{Related Work \label{sec:relwork}}

The entrenchment effect has been attracting attention for several
years \cite{sherman,power,bias,contro,googlearchy,define}, but formal
models for and analysis of the impact of search engines on the
evolution of the Web graph \cite{prefer} or on the time taken by new
pages to become popular \cite{impact} are recent.

A few solutions to the entrenchment problem have been proposed~\cite{estqual, agebased, timepage}.
They rely on variations of PageRank:
the solutions of~\cite{agebased, timepage} assign an additional weighting factor based on page age;
that of~\cite{estqual} uses the derivative of PageRank to forecast future PageRank values for young pages.

Our approach, randomized rank promotion, is quite different in spirit.
The main strength of our approach is its simplicity---it 
does not rely on measurements of the age or PageRank evolution of individual Web pages, 
which are difficult to obtain and error-prone at low sample rates.
(Ultimately, it may make sense to use our approach 
in conjunction with other techniques, in a complementary fashion.)

The exploration/exploitation tradeoff that arises in our context is akin to problems
studied in the field of reinforcement learning~\cite{rein-survey}. 
However, direct application of reinforcement
learning algorithms appears prohibitively expensive at Web scales.

\section{Model and Metrics}
\label{sec:definitions}
In this section we introduce the model of Web page popularity,
adopted from~\cite{impact}, that we use in the rest of this paper.
(For convenience, a summary of the notation we use is provided in Table~\ref{tab:symbols}.)
Recall from Section~\ref{sec:introDesign}
that in our model the Web is categorized into disjoint groups by topic, such that each page pertains
to exactly one topic. Let $\mathcal{P}$ be the set of pages devoted to a particular
topic $T$,
and let $\mathcal{U}$ denote the set of users interested in topic $T$.
Let $n = |\mathcal{P}|$ and $u = |\mathcal{U}|$ denote the number of pages and users, respectively, in the community.

\subsection{Page Popularity}
In our model, time is divided into discrete intervals,
and at the end of each interval the search engine measures
the popularity of each Web page according to
in-link count, PageRank, user traffic, or some other
indicator of popularity among users. Usually it is only possible to
measure popularity among a minority of users. 
Indeed, for in-link count or PageRank, only
those users who have the ability to create links are counted. 
For metrics based on user traffic,
typically only users who agree to install a special toolbar that monitors 
Web usage, as in~\cite{Alexa}, are counted.
Let $\mathcal{U}_m \subseteq \mathcal{U}$ denote the set of {\em monitored users}, over which page 
popularity is measured, and let $m = |\mathcal{U}_m|$. We assume $\mathcal{U}_m$ constitutes
a representative sample of the overall user population $\mathcal{U}$.

\begin{table}[t!]
\centering
\begin{tabular}{|l|p{2.45in}|}
\hline 
{\bf Symbol} & {\bf Meaning} \\ \hline\hline
$\mathcal{P}$ & Set of Web pages in community \\ \hline
$n$ & $= |\mathcal{P}|$ \\ \hline
$\mathcal{U}$ & Set of users in community \\ \hline
$u$ & $= |\mathcal{U}|$ \\ \hline
$\mathcal{U}_m$ & Set of monitored users in community \\ \hline
$m$ & $= |\mathcal{U}_m|$ \\ \hline
$P(p,t)$ & Popularity among monitored users of page $p$ at time $t$ \\ \hline
$V_u(p,t)$ & Number of user visits to page $p$ \\
         & during unit time interval at $t$ \\ \hline
$V(p,t)$ & Number of visits to $p$ by monitored users at $t$ \\ \hline
$v_u$      & Total number of user visits per unit time \\ \hline
$v$    & Number of visits by monitored users per unit time \\ \hline
$A(p,t)$ & Awareness among monitored users of page $p$ at time $t$ \\ \hline
$Q(p)$ & Intrinsic quality of page $p$ \\ \hline
$l$ & Expected page lifetime \\ \hline
\end{tabular}
\caption{Notation used in this paper.}
\vspace*{6mm}
\label{tab:symbols}
\end{table}

Let the total number of user visits to pages per unit time be fixed at $v_u$.
Further, let $v$ denote the number of visits per unit
time by monitored users, with $v = v_u \cdot \frac{m}{u}$.
The way these visits are distributed among pages in $\mathcal{P}$ is determined largely by
the search engine ranking method in use; we will come back to this aspect later.
For now we simply provide a definition of the visit rate of a page $p \in \mathcal{P}$.

\begin{definition}
(Visit Rate) The visit rate of page $p$ at time $t$, $V(p, t)$, is
defined as the number of times $p$ is visited by any monitored user 
within a unit time interval at time $t$.  
\end{definition}

\noindent
Similarly, let $V_u(p, t)$ denote the number of visits by any user in 
$\mathcal{U}$ (monitored and unmonitored users alike)
within a unit time interval at time $t$.
We require that $\forall t, \sum_{p \in \mathcal{P}} V_u(p, t) = v_u$ and 
$\forall t, \sum_{p \in \mathcal{P}} V(p, t) = v$.
Once a user visits a page for the first time, she becomes ``aware'' of that page.

\begin{definition}
(Awareness) The awareness level of page $p$ at time $t$,
$A(p,t)$, is defined as the fraction of monitored users who have visited $p$ at least once by
time $t$. 
\end{definition}

\noindent
We define the popularity of 
page $p$ at time $t$, $P(p,t) \in [0,1]$, as follows:

\begin{eqnarray}
\label{eqn:aware-to-popularity}
P(p,t) = A(p,t) \cdot Q(p)
\end{eqnarray}
where $Q(p)\in [0,1]$ ({\em page quality}) denotes the extent to which an average user 
would ``like'' page $p$ if she was aware of $p$.

In our model page popularity is a monotonically nondecreasing function of time.
Therefore if we assume nonzero page viewing probabilities, 
for a page of infinite lifetime
$\lim_{t \to \infty} P(p,t) = Q(p)$.

\subsection{Rank Promotion}

If pages are ranked strictly according to current popularity,
it can take a long time for the popularity of a new page to approach its quality.
Artificially promoting the rank of new pages can potentially
accelerate this process. One
important objective for rank promotion is to minimize the time it takes for a
new high-quality page to attain its eventual popularity, denoted {\em TBP} for ``time to become popular.''
In this paper we measure TBP as the time it takes for a high-quality page to attain popularity that
exceeds $99\%$ of its quality level.

\begin{figure}[!t]
\begin{center}
\includegraphics[width=230pt]{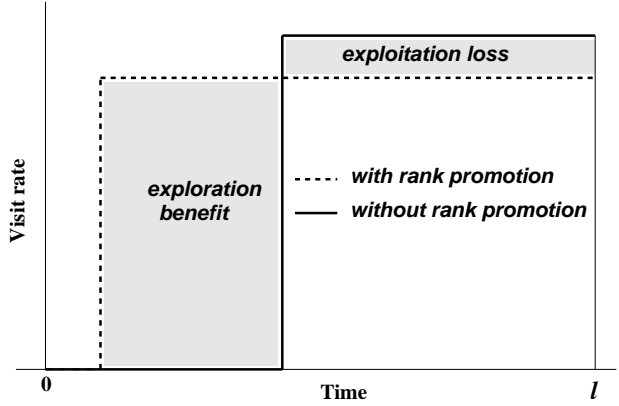}
\caption{Exploration/exploitation tradeoff.}
\label{fig:keyFig}
\end{center}
\end{figure}

Figure~\ref{fig:keyFig} shows popularity evolution curves for a particular page having very high quality
created at time $0$ with lifetime $l$, both with and without rank promotion. (It has been shown
\cite{impact} that popularity evolution curves are close to step-functions.)
Time is plotted on the x-axis. The y-axis plots the number of user visits per time unit. Note that
while the page becomes popular earlier when rank promotion is applied, the number of visits it receives
once popular is somewhat lower than in the case without rank promotion. That is because systematic
application of rank promotion inevitably comes at the cost of fewer visits to already-popular pages.

\subsection{Exploration/Exploitation Tradeoff and Quality-Per-Click Metric}

The two shaded regions of Figure~\ref{fig:keyFig} indicate the positive and negative aspects of
rank promotion. The {\em exploration benefit} area corresponds to the increase in the number of additional visits
to this particular high-quality page during its lifetime made possible by promoting it early on. 
The {\em exploitation loss} area corresponds to the decrease
in visits due to promotion of other pages, which may mostly be of low quality compared to this one.
Clearly there is a need to balance these two factors. The TBP metric is one-sided in this respect, so we
introduce a second metric that takes into account both exploitation and exploitation: 
{\em quality-per-click}, or QPC for short.
QPC measures the average quality of pages viewed by users, amortized over a long period of time.
We believe that maximizing QPC is a suitable objective for designing a rank promotion strategy.

We now derive a mathematical expression for QPC in our model.
First, recall that the number of visits by any user to page $p$ during time interval $t$ is denoted $V_u(p,t)$.
We can express the cumulative quality of all pages in $\mathcal{P}$ viewed at time $t$ as $\sum_{p \in \mathcal{P}} V_u(p,t) \cdot Q(p)$. Taking the average across time in the limit as the time duration tends to infinity, we obtain:

\[ \lim_{t \to \infty} \sum_{t_l = 0}^{t}{\sum_{p \in \mathcal{P}} \big( V_u(p,t_l) \cdot Q(p) \big)} \]
\linebreak
\linebreak \linebreak
\linebreak
\noindent
By normalizing, we arrive at our expression for QPC:

\begin{eqnarray*}
\mathit{QPC} & = & \lim_{t \to \infty} \frac{ \sum_{t_l=0}^{t}{\sum_{p \in \mathcal{P}} {\big( V_u(p,t_l) \cdot Q(p) \big)}} }
{ \sum_{t_l=0}^{t}{ \big( \sum_{p \in \mathcal{P}} V_u(p,t_l) \big) } }
\end{eqnarray*}

\section{Randomized Rank Promotion}
\label{sec:randomized}
We now describe our simple randomized rank promotion scheme 
(this description is purely conceptual; more efficient implementation techniques exist).

Let $\mathcal{P}$ denote the set of $n$ responses to a user query. A 
subset of those pages, $\mathcal{P}_p \subseteq \mathcal{P}$ is
set aside as the {\em promotion pool}, which contains the set of pages selected for rank promotion
according to a predetermined rule. (The particular rule for selecting $\mathcal{P}_p$, as well as two additional
parameters, $k \ge 1$ and $r \in [0,1]$, are configuration options that we discuss shortly.)
Pages in $\mathcal{P}_p$ are sorted randomly and the result is stored in the ordered list $\mathcal{L}_p$.
The remaining pages ($\mathcal{P} - \mathcal{P}_p$) are ranked in the usual deterministic
way, in descending order of popularity; the result is an ordered list $\mathcal{L}_d$. The two lists are merged
to create the final result list $\mathcal{L}$ according to the following procedure:

\begin{enumerate}
\item The top $k-1$ elements of $\mathcal{L}_d$ are removed from $\mathcal{L}_d$ and inserted into the beginning of $\mathcal{L}$ while preserving their order.
\item The element to insert into $\mathcal{L}$ at each remaining position $i = k, k+1, \ldots, n$ is determined one at a time, in that order, by flipping a biased coin: with probability $r$ the next element is taken from the top of list $\mathcal{L}_p$; otherwise it is taken from the top of $\mathcal{L}_d$. If one of $\mathcal{L}_p$ or $\mathcal{L}_d$ becomes empty, all remaining entries are taken from the nonempty list. At the end both of $\mathcal{L}_d$ and $\mathcal{L}_p$ will be empty, and $\mathcal{L}$ will contain one entry for each of the $n$ pages in $\mathcal{P}$.
\end{enumerate}

\noindent
The configuration parameters are:

\begin{itemize}
\item {\bf Promotion pool ($\mathcal{P}_p$):} In this paper we consider two rules for determining which pages
are promoted: (a) the {\em uniform} promotion rule, in which every page is included in $\mathcal{P}_p$ with equal probability $r$,
and (b) the {\em selective} promotion rule, in which all pages whose current awareness level among monitored users is zero (i.e., $\mathcal{A}(p,t) = 0$) are included in $\mathcal{P}_p$, and no others. (Other rules are of course possible; we chose to focus on these two in particular because they roughly correspond to the extrema of the spectrum of interesting rules.)
\item {\bf Starting point ($k$):} All pages whose natural rank is better than $k$ are protected from the effects of promoting other pages. A particularly interesting value is $k=2$, which safeguards the top result of any search query, thereby preserving the ``feeling lucky'' property that is of significant value in some situations.
\item {\bf Degree of randomization ($r$):} When $k$ is small, this parameter governs the tradeoff between emphasizing exploration (large $r$) and emphasizing exploitation (small $r$).
\end{itemize}

\noindent

Our goal is to determine settings of the above parameters that lead to good TBP and QPC values. The remainder of
this paper is dedicated to this task. 
Next we present our analytical model of Web page popularity evolution, which we use to estimate TBP and QPC under various ranking methods.

\section{Analytical Model}
\label{sec:TBP}

Our analytical model has these features:
\begin{itemize} \itemsep0pt
  \item Pages have finite lifetime following an exponential distribution
    (Section~\ref{sec-lifetime}).  The number of pages and the number
    of users are fixed in steady state.  The quality distribution of
    pages is stationary.
  \item The expected awareness, popularity, rank, and visit rate of a page
    are coupled to each other through a combination of the search engine
    ranking function and the bias in user attention to search results
    (Sections~\ref{sec:aware} and~\ref{sec:g}).
\end{itemize}

Given that (a)~modern search engines appear to be strongly influenced by
popularity-based measures while ranking results, and
(b)~users tend to focus their attention primarily on the top-ranked 
results~\cite{joachims,lempel}, it is reasonable to assume
that the expected visit rate of a page is a function of its current popularity 
(as done in~\cite{impact}):
\begin{eqnarray}
\label{eqn:ran-vtop-mapping}
V(p,t) &=& F(P(p,t))
\end{eqnarray} 
where the form of function $F(x)$ depends on the ranking method in use and the
bias in user attention. For example, if ranking
is completely random, then $V(p,t)$ is independent
of $P(p,t)$ and the same for all pages, so
$F(x) = v \cdot \frac{1}{n}$.  (Recall that $v$ is the total number
of monitored user visits per unit time.)  If ranking is done
in such a way that user traffic to a page
is proportional to the popularity of that page,
$F(x) = v \cdot \frac{x}{\phi}$, where $\phi$ is a normalization
factor; at steady-state, 
$\phi = \sum_{p \in \mathcal{P}} P(p,t)$.
If ranking is performed the aforementioned way $50\%$ of the time,
and performed randomly $50\%$ of the time, then
$F(x) = v \cdot \big( 0.5 \cdot \frac{x}{\phi} + 0.5 \cdot \frac{1}{n} \big)$.
For the randomized rank promotion we introduced in Section~\ref{sec:randomized}
the situation is more complex.  We defer discussion of how
to obtain $F(x)$ to Section~\ref{sec:g}.

\subsection{Page Birth and Death \label{sec-lifetime}}

The set of pages on the Web is not fixed. Likewise, we assume that for
a given community based around topic $T$, the set $\mathcal{P}$
of pages in the community evolves over time due to pages being created 
and retired. To keep our analysis manageable we assume that the rate
of retirement matches the rate of creation, so that the total number
of pages remains fixed at $n = |\mathcal{P}|$. We model retirement
of pages as a Poisson
process with rate parameter $\lambda$, so the expected lifetime of a 
page is $l = \frac{1}{\lambda}$ 
(all pages have the same expected lifetime\footnote{In reality, 
lifetime might be a positively correlated with popularity. 
If so, popular pages would remain entrenched for a longer time than under our model,
leading to even worse TBP than our model predicts.}).
When a page is retired, 
a new page of equal quality is created immediately, so the distribution of page quality
values is stationary.  When a new page is created it has initial
awareness and popularity values of zero.

\subsection{Awareness Distribution \label{sec:aware}}

We derive an expression for the distribution of page awareness values,
which we then use to obtain an expression for quality-per-click (QPC).
We analyze the steady-state scenario, in which the awareness and popularity
distributions have stabilized and remain steady over time.
Our model may not seem to indicate steady-state behavior, 
because the set of pages is constantly in flux
and the awareness and popularity of an individual page 
changes over time. To understand the basis for 
assuming steady-state behavior, consider the set $\mathcal{C}_t$ of pages created at time $t$,
and the set $\mathcal{C}_{t+1}$ of pages created at time $t+1$. Since
page creation is governed by a Poisson process the expected sizes of the two sets are equal. 
Recall that we assume the distribution of page quality values remains the same
at all times. 
Therefore, the popularity of all pages in both $\mathcal{C}_t$ and $\mathcal{C}_{t+1}$ will
increase from the starting value of $0$ according to the same popularity evolution law.
At time $t+1$, when the pages in $\mathcal{C}_t$ have evolved in popularity according
to the law for the first time unit, the new pages in $\mathcal{C}_{t+1}$ introduced at time $t+1$
will replace the old popularity values of the $\mathcal{C}_t$ pages. A symmetric effect
occurs with pages that are retired, resulting in steady-state behavior overall. 
In the steady-state, both popularity and awareness distributions are stationary.

The steady-state awareness distribution is given as follows.

\begin{theorem}\label{th:awareness}
  Among all pages in $\mathcal{P}$ whose quality is $q$, the fraction
  that have awareness $a_i = \frac{i}{m}$ (for $i = 0, 1, \dots, m$) is:
  \begin{equation}
    \label{eqn:recursion5}
    f(a_{i}|q) = \frac{\lambda}{(\lambda + F(0)) \cdot (1-a_{i})} \prod_{j=1}^{i} 
    \frac{F(a_{j-1} \cdot q)}{\lambda + F(a_{j} \cdot q)}
  \end{equation}
  where $F(x)$ is the function in Equation~\ref{eqn:ran-vtop-mapping}.
\end{theorem}
\begin{proof}
See Appendix~\ref{apx:thm1Proof}.
\end{proof}

\begin{figure*}[!t]
\begin{center}
\includegraphics[width=200pt]{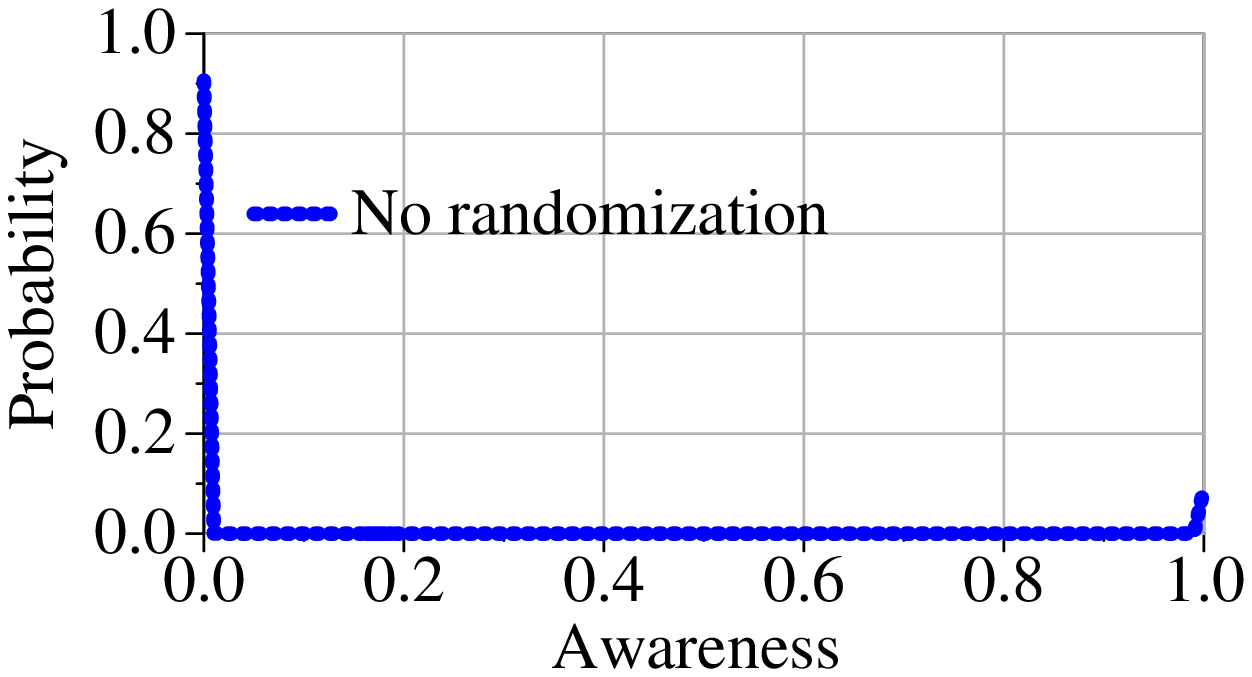} 
\hspace*{20mm}\includegraphics[width=200pt]{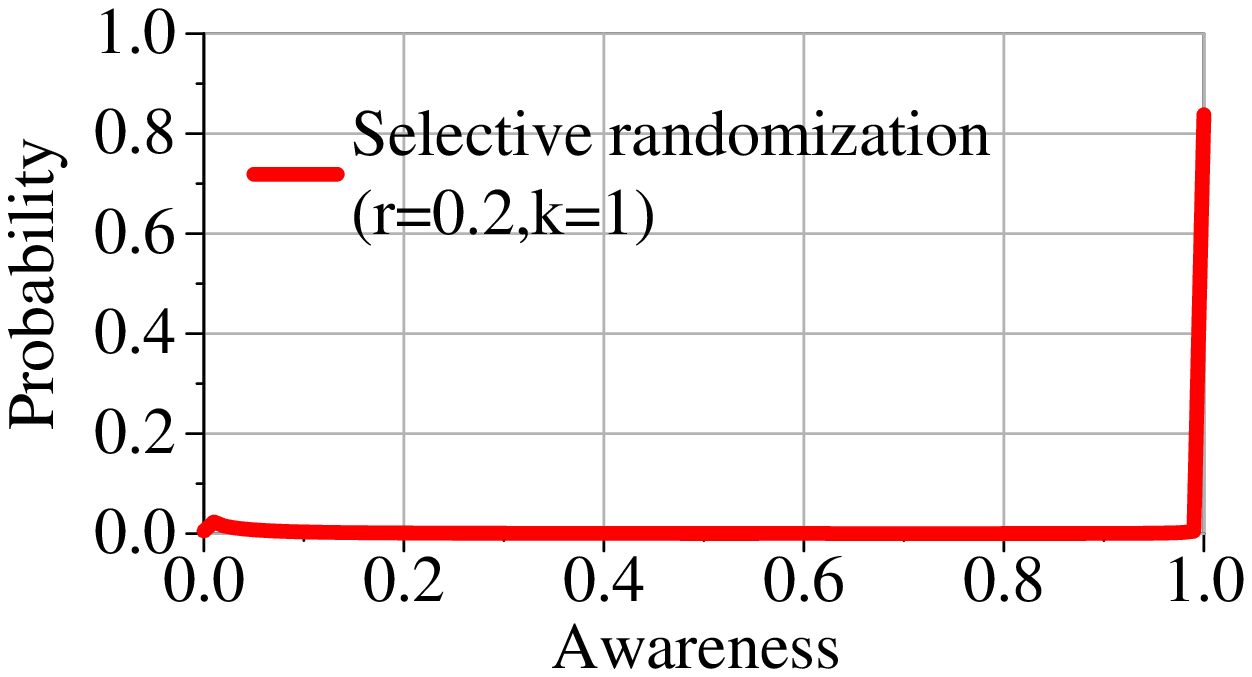}
\caption{Awareness distribution of pages of high quality under randomized and nonrandomized ranking.}
\label{fig:awareness}
\end{center}
\end{figure*}

Figure~\ref{fig:awareness} plots the steady-state awareness distribution for pages of highest quality,
under both nonrandomized ranking and selective randomized rank promotion with $k=1$ and $r=0.2$,
for our default Web community characteristics (see Section~\ref{sec:defaultSettings}).
For this graph we used the procedure described in Section~\ref{sec:g} to obtain the function $F(x)$. 

Observe that if randomized rank promotion is used, in steady-state most
high-quality pages have large awareness, whereas if standard
nonrandomized ranking is used most pages have very small awareness.
Hence, under randomized
rank promotion most pages having high quality spend most of their lifetimes
with near-$100\%$ awareness, yet with nonrandomized ranking they
spend most of their lifetimes with near-zero awareness. 
Under either ranking scheme pages spend very little time in the middle
of the awareness scale, since the rise to high awareness is nearly a
step function.

Given an awareness distribution $f(a|q)$, it is straightforward to determine expected time-to-become-popular (TBP) 
corresponding to a given quality value (formula omitted for brevity).
Expected quality-per-click (QPC) is expressed as follows:

  \begin{equation*}
    QPC = \frac{
      \sum_{p \in \mathcal{P}} \sum_{i=0}^{m} f(a_i|Q(p)) \cdot F(a_i \cdot Q(p)) \cdot Q(p)
      }{
      \sum_{p \in \mathcal{P}} \sum_{i=0}^{m} f(a_i|Q(p)) \cdot F(a_i \cdot Q(p)) 
    }
  \end{equation*}

\noindent
where $a_i = \frac{i}{m}$. (Recall our assumption that monitored users are a representative sample of all users.)

\subsection{Popularity to Visit Rate Relationship \label{sec:g}}

In this section we derive the function $F(x)$ used in
Equation~\ref{eqn:ran-vtop-mapping}, which governs the 
relationship between $P(p,t)$ and the expectation of $V(p,t)$.
As done in~\cite{impact} we
split the relationship between
the popularity of a page and the expected number of visits into two components: 
(1)~the relationship between popularity and rank position, and 
(2)~the relationship between rank position and the number of visits.
We denote these two relationships as the functions $F_1$ and $F_2$ respectively, 
and write:
\begin{eqnarray*}
F(x) &=& F_2(F_1(x))
\end{eqnarray*}
where the output of $F_1$ is the rank position of a page of popularity $x$,
and $F_2$ is a function from that rank to a visit rate.
Our rationale for splitting $F$ in this way is that, according
to empirical findings reported in~\cite{joachims}, the likelihood of a user visiting
a page presented in a search result list depends primarily on the rank position at which the page appears.

We begin with $F_2$, the dependence of the expected number of user visits
on the rank of a page in a result list.  Analysis of
AltaVista usage logs \cite{impact,lempel} reveal that the following relationship holds 
quite closely\footnote{User views were measured at the granularity of groups of ten 
results in~\cite{lempel}, and later extrapolated to individual pages in~\cite{impact}.}:

\begin{eqnarray}
\label{eqn:rtop-mapping}
F_2(x) &=& \theta \cdot x^{-3/2}
\end{eqnarray}

\noindent
where $\theta$ is a normalization constant, which we set as:
\begin{eqnarray*}
\theta &=& \frac{v}{\sum_{i=1}^n i^{-3/2}}
\end{eqnarray*}
where $v$ is the total number of monitored user visits per unit time.

Next we turn to $F_1$, the dependence of rank on the popularity of a page.
Note that since the awareness level of a particular page cannot be pinpointed
precisely (it is expressed as a probability distribution), we express $F_1(x)$
as the {\em expected} rank position of a page of popularity $x$. In doing so
we compromise accuracy to some extent, since we will determine the expected
number of visits by applying $F_2$ to the expected rank, as opposed to summing over
the full distribution of rank values. (We examine the accuracy of our analysis in
Sections~\ref{sec:analysisValidation} and~\ref{sec:analysisValidation2}.)

Under nonrandomized ranking, the expected rank of a page of popularity $x$ is
one plus the expected
number of pages whose popularities surpass $x$.
By Equation~\ref{eqn:aware-to-popularity}, page $p$ has $P(p,t)>x$ if it has
$A(p,t) > x/Q(p)$. From Theorem~\ref{th:awareness} the probability that a randomly-chosen
page $p$ satisfies this condition is:

\[ \sum_{i = 1 + \lfloor m \cdot x/Q(p) \rfloor }^{m} f \left(\left. \frac{i}{m} \right| Q(p) \right) \]

\noindent
By linearity of expectation,
summing over all $p \in \mathcal{P}$ we arrive at:

  \begin{equation}
    \label{eq:popularity}
    F_1(x) \approx 1 + \sum_{p \in \mathcal{P}}
    \left(
      \sum_{i = 1 + \lfloor m \cdot x/Q(p) \rfloor }^{m} f\left(\left.\frac{i}{m}\right|Q(p) \right)
    \right)
  \end{equation}

\noindent
(This is an approximate expression because we ignore the effect of ties in popularity values, and because
we neglect to discount one page of popularity $x$ from the outer summation.)

The formula for $F_1$ under uniform randomized ranking is rather complex,
so we omit it. We focus instead on selective randomized ranking,
which is a more effective strategy, as we will demonstrate shortly.
Under selective randomized ranking the expected rank of a page of popularity
$x$, when $x > 0$, is given by:

  \begin{displaymath}
    \label{eq:sel-popularity}
    F'_1(x) \approx \left\{ \begin{array}{ll}
      F_1(x) & \textrm{if $F_1(x) < k$} \\
      F_1(x) + \min \{\frac{r \cdot (F_1(x)-k+1)}{(1-r)}, z\} & \textrm{otherwise}
      \end{array}
    \right.
  \end{displaymath}
where $F_1$ is as in Equation~\ref{eq:popularity},
and $z$ denotes the expected number of pages with zero awareness, an estimate for
which can be computed without difficulty under our steady-state assumption.
(The case of $x = 0$ must be handled separately;
we omit the details due to lack of space.)

The above expressions for $F_1(x)$ or $F'_1(x)$ each contain a circularity,
because our formula for $f(a|q)$ (Equation~\ref{eqn:recursion5}) contains
$F(x)$. It appears that a closed-form solution for $F(x)$
is difficult to obtain. In the absence of a closed-form
expression one option is to determine $F(x)$ via simulation.
The method we use is to solve for $F(x)$ using
an iterative procedure, as follows.

We start with a simple function for $F(x)$, say $F(x) = x$,
as an initial guess at the solution. We then substitute this
function into the right-hand side of the appropriate equation
above to produce a new $F(x)$ function in numerical form.
We then convert the numerical $F(x)$ function into symbolic form by fitting a curve,
and repeat until convergence occurs.
(Upon each iteration we adjust the curve slightly so as
to fit the extreme points corresponding to $x = 0$ and $x = 1$ especially carefully; details
omitted for brevity.)
Interestingly, we found that using a quadratic curve in log-log space led to good
convergence for all parameter settings we tested, so that:

\begin{equation*}
  \log F =
  \alpha \cdot (\log x)^2  + \beta \cdot \log x + \gamma
\end{equation*}

\noindent
where $\alpha$, $\beta$, and $\gamma$ are determined using a curve fitting procedure.
We later verified
via simulation that across a variety of scenarios $F(x)$ can be fit quite accurately
to a quadratic curve in log-log space.

\section{Effect of Randomized Rank Promotion and Recommended Parameter Settings}
\label{sec:experiments}

In this section we report our measurements of the impact of randomized rank promotion on search engine quality.
We begin by describing the default Web community scenario we use in Section~\ref{sec:defaultSettings}.
Then we report the effect of randomized rank promotion on TBP and QPC in Sections~\ref{sec:analysisValidation}
and~\ref{sec:analysisValidation2}, respectively. 
Lastly, in Section~\ref{sec:efConfigs} we investigate
how to balance exploration and exploitation, and give our recommended recipe for randomized rank promotion.

\begin{figure}[htbp]
  \begin{center}
    \mbox{
      \subfigure[Popularity evolution of a page of quality $Q = 0.4$ under nonrandomized, uniform randomized, and selective randomized ranking.]{{\includegraphics[width=180pt]{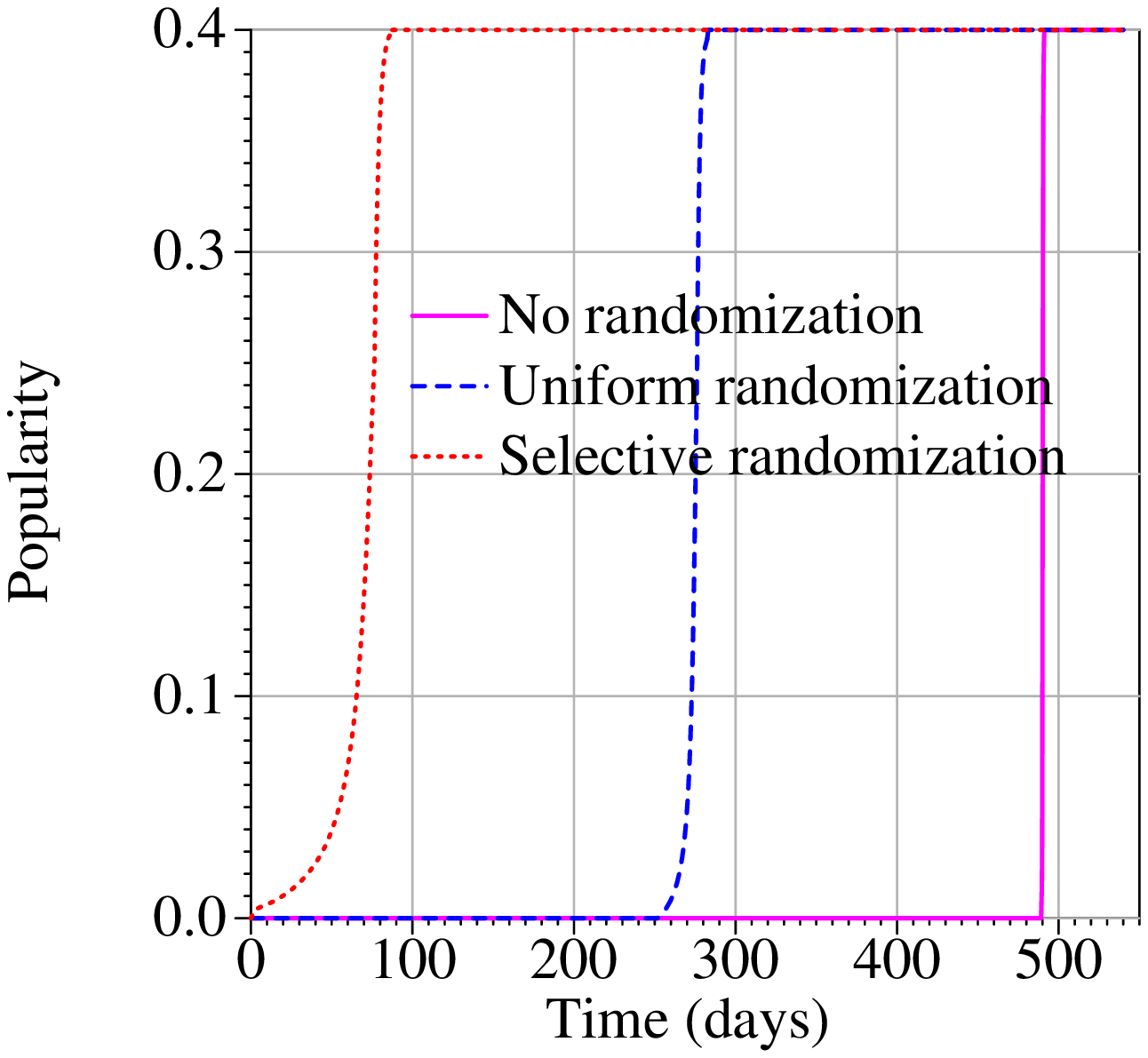}} \label{fig:onepopevolana}} \quad \quad \quad \quad \quad \quad
      \subfigure[Time to become popular (TBP) for a page of quality $0.4$ in default Web community as degree of randomization ($r$) is varied.]{{\includegraphics[width=180pt]{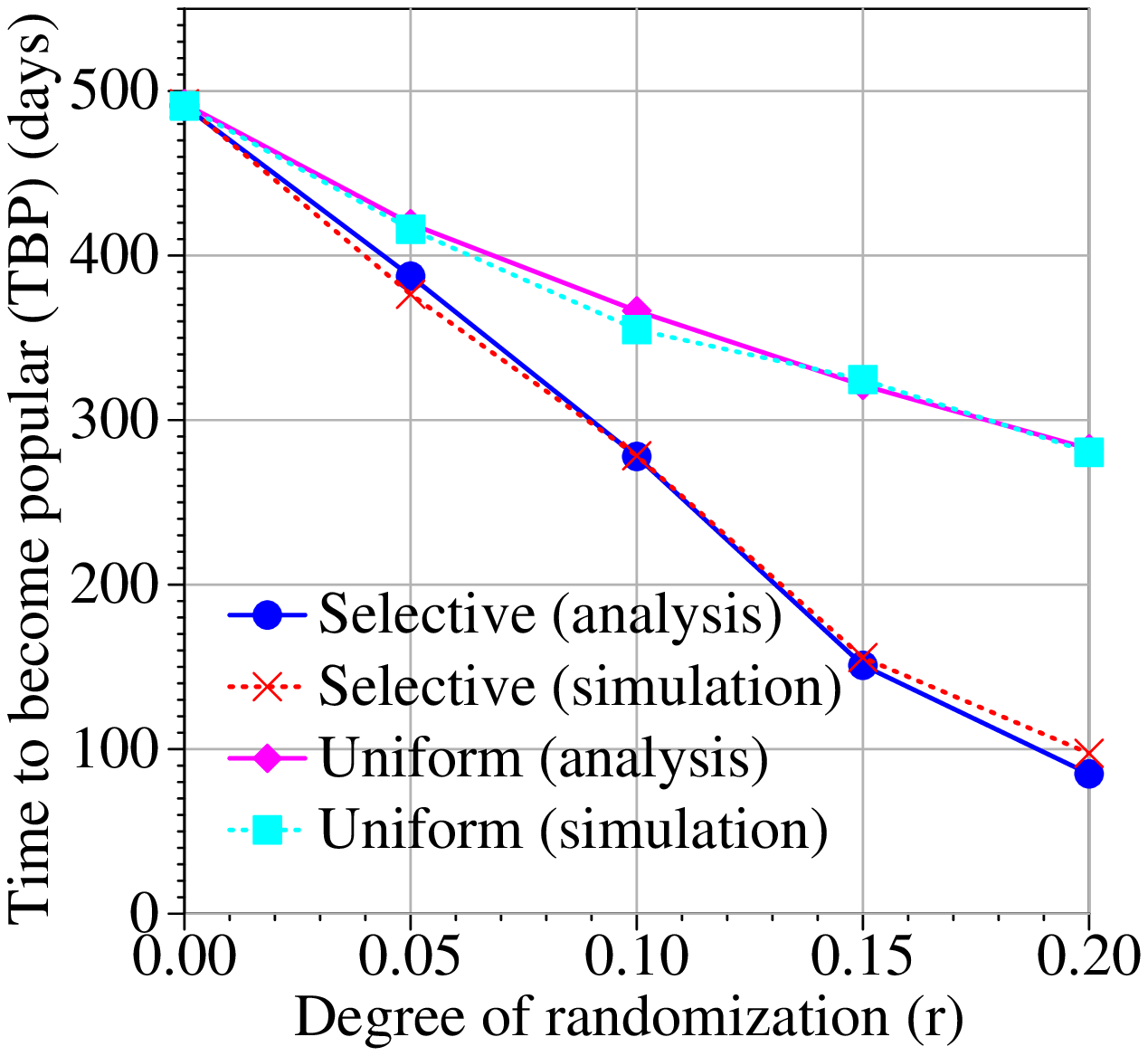}} \label{fig:tbpanasim}}
      }
    \caption{Effect of randomized rank promotion on TBP.}
  \end{center}
\end{figure}

\subsection{Default Scenario}
\label{sec:defaultSettings}

For the results we report in this paper, the 
default\footnote{We supply results for other community types in Section~\ref{sec:robustness}.} 
Web community we use is one having $n = 10,000$ pages. 
The remaining characteristics of our default Web community are set so as to be
in proportion to observed characteristics of the entire Web, as follows.
First, we set the expected page lifetime to $l = 1.5$ years (based on data from~\cite{new}).
Our default Web community has $u = 1000$ users making a total of $v_u = 1000$ visits per day 
(based on data reported in~\cite{SIMS}, the number of Web users is roughly one-tenth the number of pages, 
and an average user queries a search engine about once per day). We assume that a search engine
is able to monitor $10\%$ of its users, so $m = 100$ and $v = 100$.

As for page quality values, we had little basis for measuring the intrinsic quality distribution of pages on the Web.
As the best available approximation, we used the power-law distribution reported for
PageRank in~\cite{impact},
with the quality value of the highest-quality page set to $0.4$.
(We chose $0.4$ based on the fraction of Internet users
who frequent the most popular Web portal site, according to~\cite{searchEngineWatch}.)

\subsection{Effect of Randomized Rank Promotion on TBP}
\label{sec:analysisValidation}

Figure~\ref{fig:onepopevolana} shows popularity evolution curves derived from the awareness distribution
determined analytically for a page of quality $0.4$
under three different ranking methods: (1) nonrandomized ranking,
(2) randomized ranking using uniform promotion with the starting 
point $k=1$ and the degree of randomization $r=0.2$, and (3) randomized ranking using
selective promotion with $k=1$ and $r=0.2$. This graph shows that, not surprisingly, 
randomized rank promotion can improve TBP by a large margin.
More interestingly it also indicates that selective rank promotion achieves substantially better TBP than uniform promotion.
Because, for small $r$, there is limited opportunity to promote pages, focusing on pages with zero awareness
turns out to be the most effective method. 

Figure~\ref{fig:tbpanasim} shows TBP measurements  
for a page of quality $0.4$ in our default Web community, 
for different values of $r$ (fixing $k=1$). 
As expected, increased randomization leads to lower TBP, especially if
selective promotion is employed.

To validate our analytical model, we 
created a simulator that maintains an evolving ranked list of pages (the ranking method used
is configurable), and distributes user visits to pages according to 
Equation~\ref{eqn:rtop-mapping}. Our simulator
keeps track of awareness and popularity values of individual pages
as they evolve over time, and creates and retires pages as dictated by our model.
After a sufficient period of time has passed to reach steady-state behavior, we take measurements.

These results are plotted in Figure~\ref{fig:tbpanasim}, side-by-side with our analytical results.
We observe a close correspondence between our analytical model and our simulation.\footnote{Our analysis is 
only intended to be accurate for small values of $r$, which is why we only plot results for $r < 0.2$.
From a practical standpoint only small values of $r$ are of interest.}

\subsection{Effect of Randomized Rank Promotion on QPC}
\label{sec:analysisValidation2}

\begin{figure}[t]
\begin{center}
\includegraphics[width=200pt]{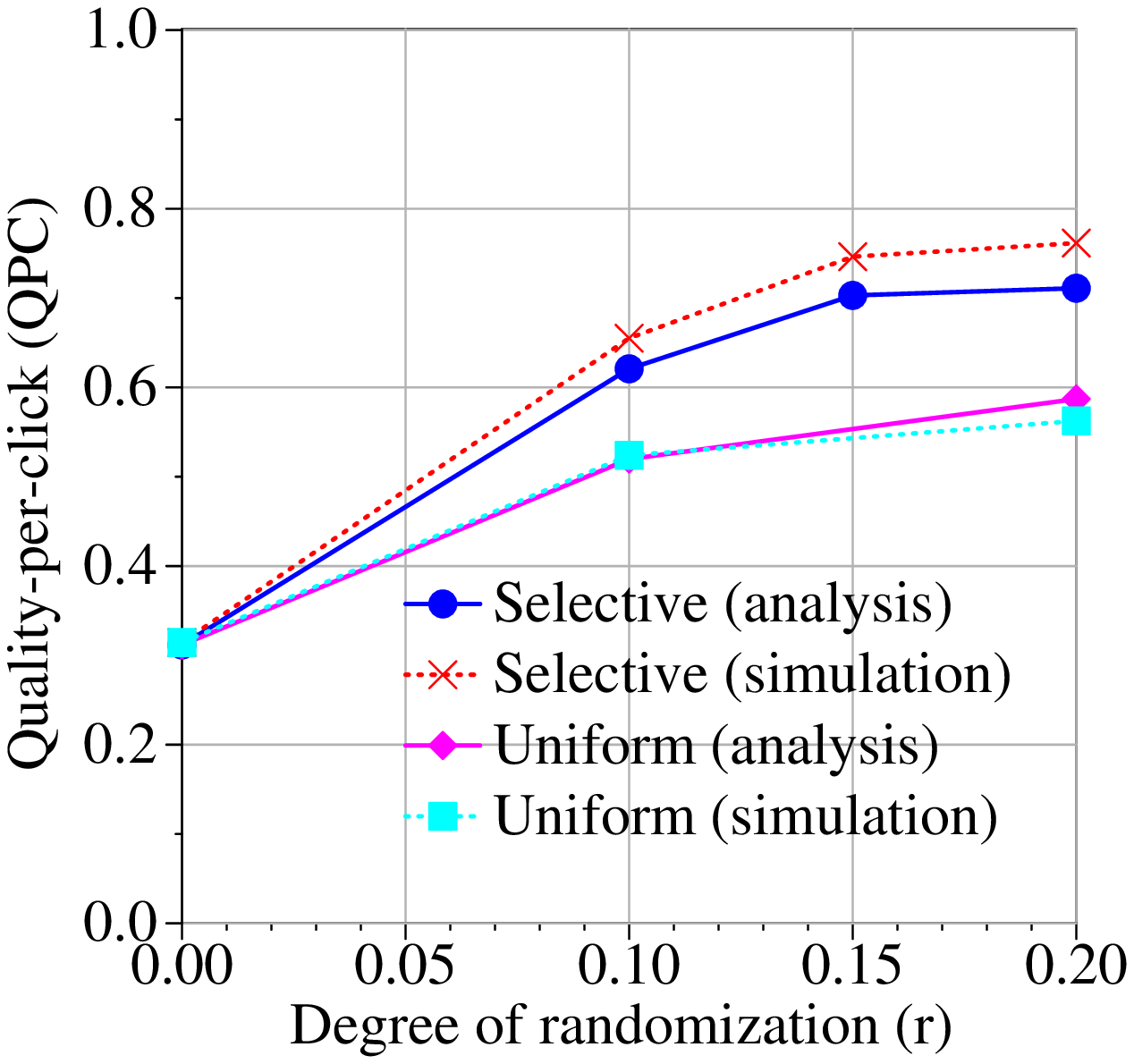}
\vspace{-6mm}
\caption{Quality-per-click (QPC) for default Web community as degree of randomization ($r$) is varied.}
\label{fig:qpcanasim}
\end{center}
\end{figure}

We now turn to quality-per-click (QPC). Throughout this paper (except in Section~\ref{sec:mixedbrowsing}) 
we normalize all QPC measurements such that
$QPC = 1.0$ corresponds to the theoretical upper bound achieved by ranking pages in descending order of quality.
The graph in Figure~\ref{fig:qpcanasim} plots normalized QPC as we vary the promotion rule and the 
degree of randomization $r$ (holding $k$ fixed at $k=1$), under our
default Web community characteristics of Section~\ref{sec:defaultSettings}.
For a community with these characteristics, a moderate dose of randomized rank promotion
 increases QPC substantially, especially
under selective promotion.

\subsection{Balancing Exploration, Exploitation, and Reality}
\label{sec:efConfigs}

We have established a strong case that 
selective rank promotion is superior to 
uniform promotion.
In this section we investigate how to set the other two randomized rank promotion parameters, $k$ and $r$, so as 
to balance exploration and exploitation and achieve high QPC.
For this purpose we prefer to rely on simulation, as opposed to analysis, for maximum accuracy.

\begin{figure}[!t]
\begin{center}
\includegraphics[width=200pt]{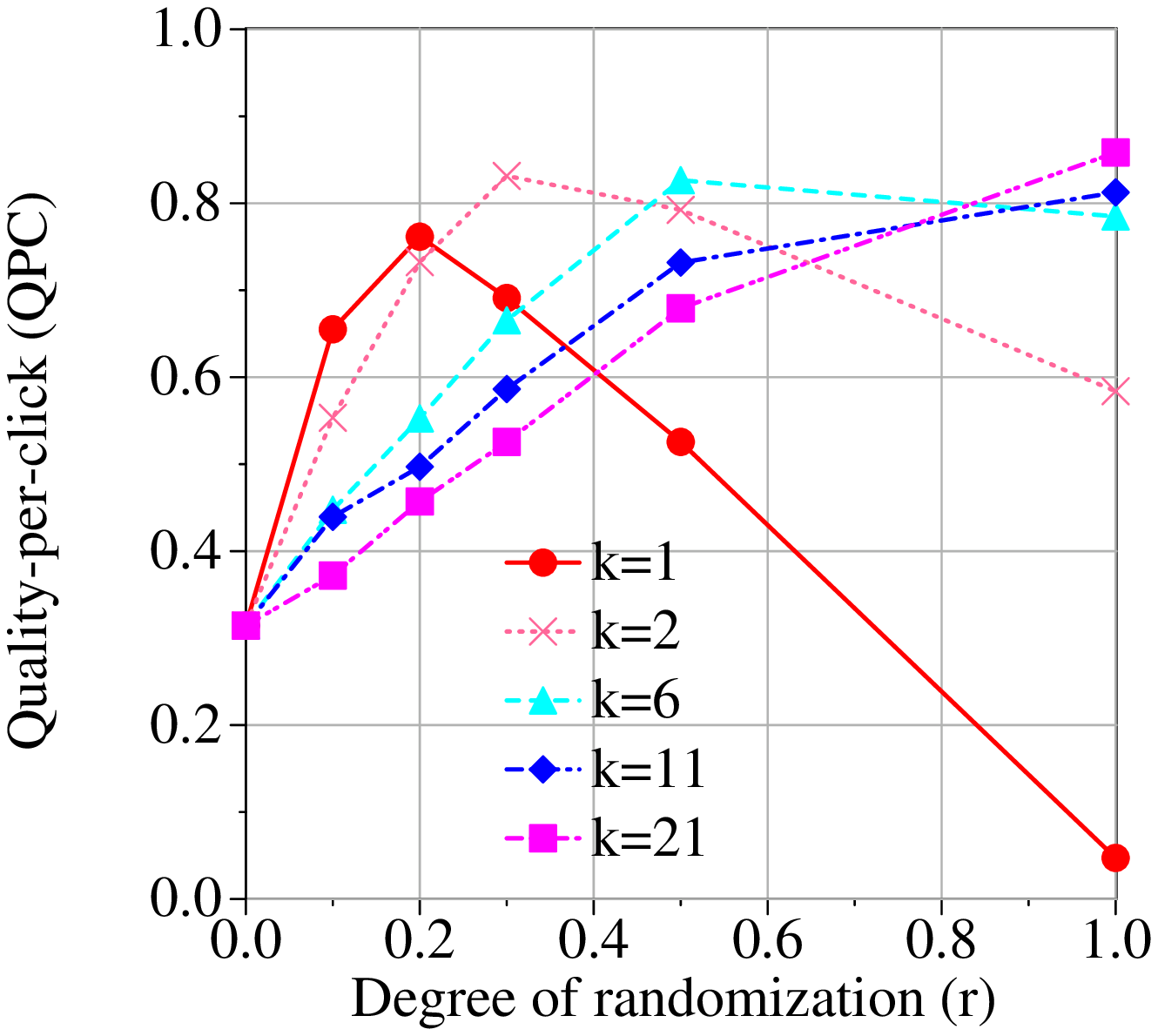}
\vspace{-6mm}
\caption{Qualitiy-per-click (QPC) for default Web community under selective 
randomized rank promotion, as degree of randomization ($r$) and starting point ($k$) are varied.}
\label{fig:kandr}
\end{center}
\end{figure}

The graph in Figure~\ref{fig:kandr} plots normalized QPC as we vary 
both $k$ and $r$, under our default scenario (Section~\ref{sec:defaultSettings}).
As $k$ grows larger, a higher $r$ value is needed to achieve high 
QPC. Intuitively, as the starting point for rank promotion becomes 
lower in the ranked list (larger $k$), a denser concentration of 
promoted pages (larger $r$) is required to ensure that new high-quality 
pages are discovered by users. 

For search engines, we take the view that
it is undesirable to include a noticeable amount of randomization in ranking, regardless of the starting point $k$. 
Based on Figure~\ref{fig:kandr}, using only $10\%$ randomization ($r=0.1$) appears
sufficient to achieve most of the benefit of rank promotion, as long as $k$ is kept small (e.g., $k=1$ or $2$). 
Under $10\%$ randomization, roughly one page in every group of ten query results 
is a new, untested page, as opposed to an established page.
We do not believe most users are likely to notice this effect, given the 
amount of noise normally present in search engine results.

A possible exception is for the topmost query result, which users often expect to be consistent if they issue the same query
multiple times. Plus, for certain queries users expect to see a single, ``correct,'' answer in the top rank position
(e.g., most users would expect the query ``Carnegie Mellon'' to return a link to the Carnegie Mellon University home page 
at position $1$), and quite a bit of effort goes into ensuring that
search engines return that result at the topmost rank position. That is why we include the $k=2$ parameter setting, which
ensures that the top-ranked search result is never perturbed.

\vspace{9pt}
\noindent
{\bf Recommendation:} {\em Introduce $10\%$ randomization starting at rank position $1$ or $2$, 
and exclusively target zero-awareness pages for random rank promotion.}

\section{Robustness Across Different Community Types}
\label{sec:robustness}

In this section we investigate the robustness of our recommended ranking method
(selective promotion rule, $r=0.1$, $k \in \{ 1, 2 \}$)
as we vary the characteristics of our testbed Web community.
Our objectives are to demonstrate: (1) that if we consider a wide range of community types, 
amortized search result quality is never harmed by our randomized rank promotion scheme,
and (2) that our method improves result quality substantially in most cases, 
compared with traditional deterministic ranking.
In this section we rely on simulation rather than analysis to ensure maximum accuracy.

\vspace{-3mm}

\subsection{Influence of Community Size}

\begin{figure}[t!]
  \begin{center}
    \mbox{
      \subfigure[Influence of community size.]{{\includegraphics[width=200pt]{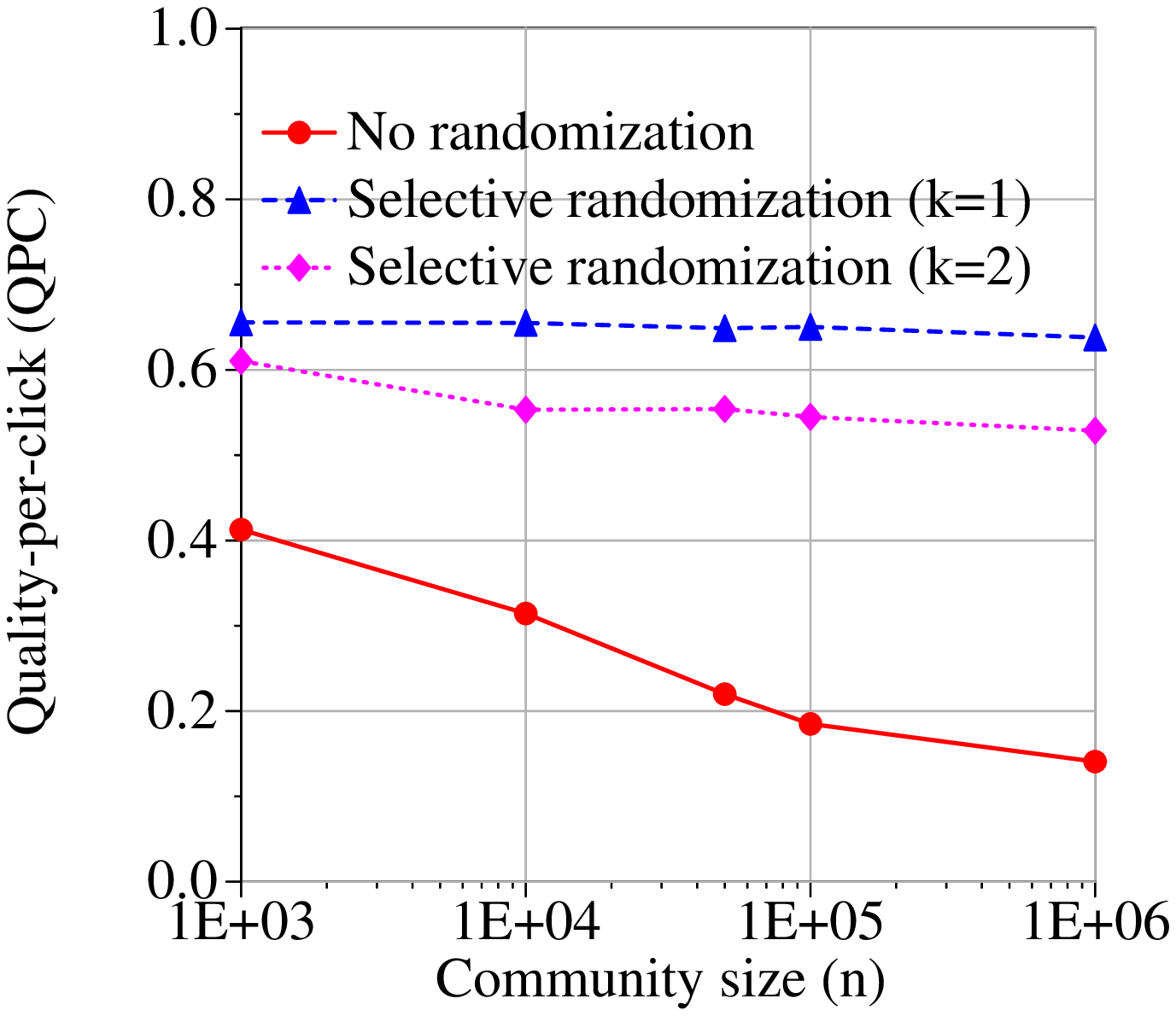}} \label{fig:page}} \quad \quad \quad \quad \quad \quad
      \subfigure[Influence of page lifetime.]{{\includegraphics[width=200pt]{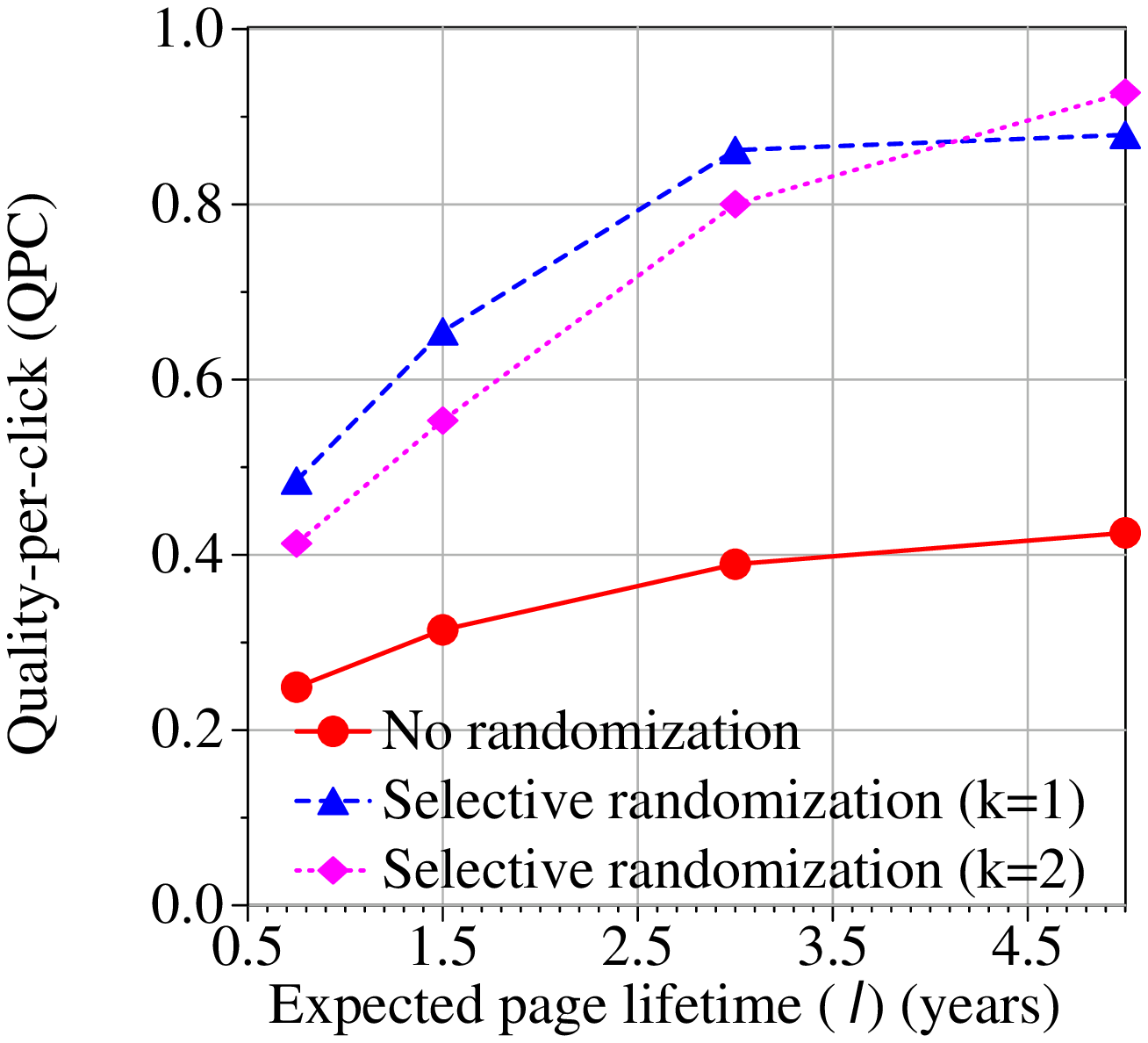}} \label{fig:life}} 
      }
    \mbox{
      \subfigure[Influence of visit rate.]{{\includegraphics[width=200pt]{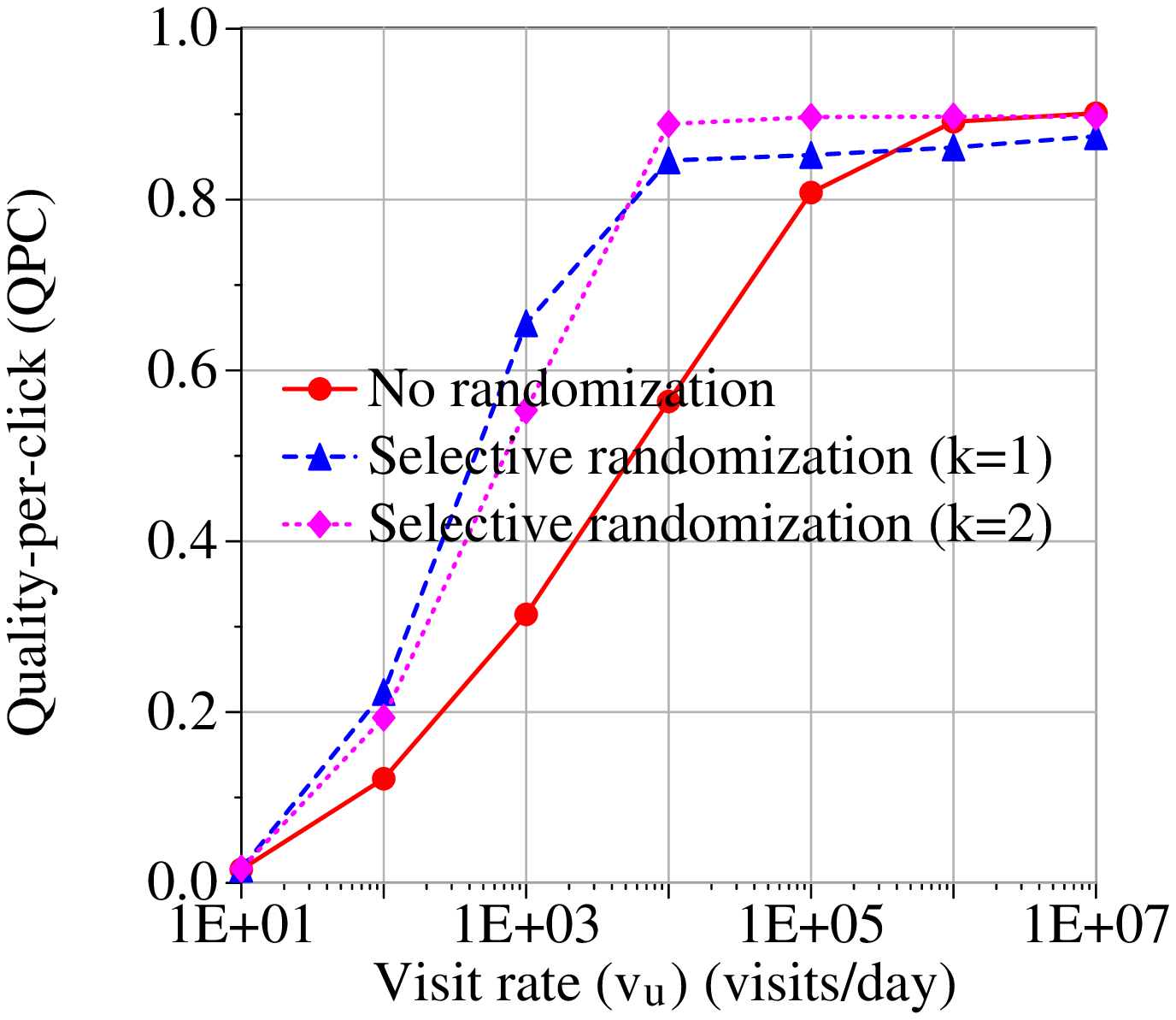}} \label{fig:view}} \quad \quad \quad \quad \quad \quad
      \subfigure[Influence of size of user population.]{{\includegraphics[width=200pt]{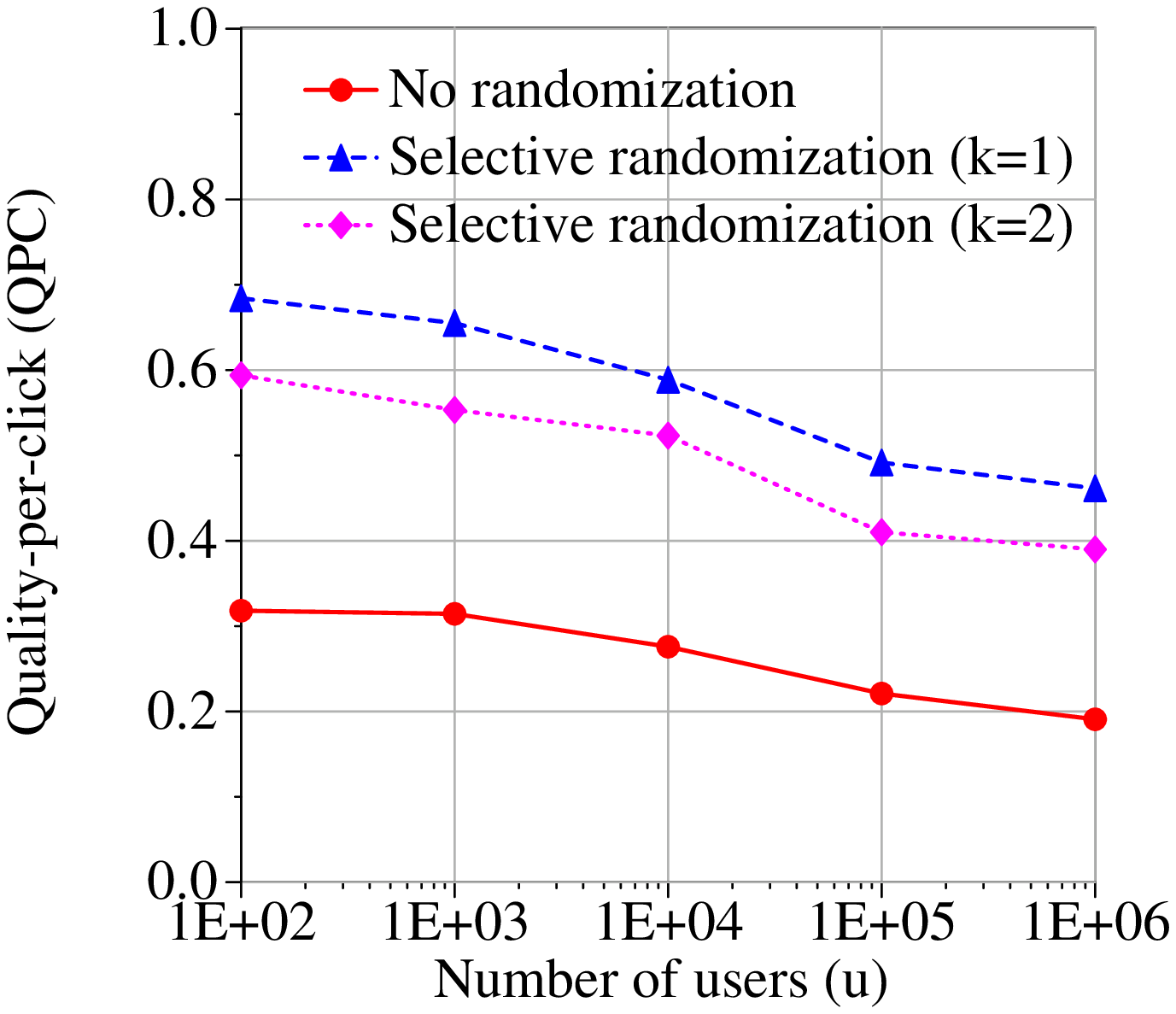}} \label{fig:user}} 
      }
    \caption{Robustness across different community types.}
  \end{center}
\end{figure}

Here we vary the number of pages in the community, $n$, while holding the ratio of users to pages fixed at $u/n = 10\%$,
fixing the fraction of monitored users as $m/u = 10\%$, and
fixing the number of daily page visits per user at $v_u/u = v/m = 1$.
Figure~\ref{fig:page} shows the result, with community size $n$ plotted on the x-axis on a logarithmic scale.
The y-axis plots normalized QPC for three different ranking methods: nonrandomized, selective randomized with $r=0.1$ and $k=1$,
and selective randomized with $r=0.1$ and $k=2$.
With nonrandomized ranking, QPC declines as community size increases, because
it becomes more difficult for new high-quality pages to overcome the entrenchment effect.
Under randomized rank promotion, on the other hand, 
due to rank promotion QPC remains 
high and fairly steady across a range of community sizes.

\vspace{-2mm}

\subsection{Influence of Page Lifetime}
\label{sec:ans-life}

Figure~\ref{fig:life} shows QPC as we vary the expected page 
lifetime $l$ while keeping all other community characteristics fixed.
(Recall that in our model the number of pages in the community remains 
constant across time, and when a page is retired a new one
of equal quality but zero awareness takes its place.)
The QPC curve for nonrandomized ranking confirms our intuition:
when there is less churn in the set of pages in the community (large $l$),
QPC is penalized less by the entrenchment effect. More interestingly, 
the margin of improvement in QPC over nonrandomized ranking due to
introducing randomness is greater when pages
tend to live longer. The reason is that with a low page creation rate 
the promotion pool can be kept small. Consequently new pages benefit from
larger and more frequent rank boosts, on the whole, helping the high-quality
ones get discovered quickly.

\subsection{Influence of Visit Rate}

The influence of the aggregate user visit rate on QPC is plotted in Figure~\ref{fig:view}.
Visit rate is plotted on the x-axis on a logarithmic scale, and QPC is plotted on the y-axis.
Here, we hold the number of pages fixed at our default value of $n=10,000$ 
and use our default expected lifetime value of $l=1.5$ years. We vary
the total number of user visits per day $v_u$ while holding the ratio of daily page visits to users
fixed at $v_u/u = 1$ and, as always, fixing the fraction of monitored users as $m/u = 10\%$.
From Figure~\ref{fig:view} we see first of all 
that, not surprisingly, popularity-based ranking fundamentally fails if very few
pages are visited by users. Second, if the number of visits is very large 
($1000$ visits per day to an average page), then there is no need for randomization 
in ranking (although it does not hurt much). For visit rates within an order of magnitude
on either side of $0.1 \cdot n = 1000$, which matches the average visit rate of search engines in 
general when $n$ is scaled to the 
size of the entire Web,~\footnote{According to our rough estimate based on data from~\cite{SIMS}.}
there is significant benefit to using randomized rank promotion.

\subsection{Influence of Size of User Population}

Lastly we study the affect of varying the number of users in the community $u$, while holding all other parameters
fixed: $n=10,000$, $l=1.5$ years, $v_u = 1000$ visits per day, and $m/u = 10\%$.
Note that we keep the total number of visits per day fixed, but vary the number of users making those visits.
The idea is to compare communities in which most page visits come from a core group of fairly active users
to ones receiving a large number of occasional visitors. Figure~\ref{fig:user} shows the result,
with the number of users $u$ plotted on the x-axis on a logarithmic scale, and QPC plotted on the y-axis.
All three ranking methods perform somewhat worse when the pool of users is large, although the performance
ratios remain about the same. The reason for this trend is that with a larger user pool, a stray visit to a new high-quality page
provides less traction in terms of overall awareness.

\section{Mixed Surfing and Searching \label{sec:mixedbrowsing}}
The model we have explored thus far assumes
that users make visit to pages only by querying a search engine.
While a very large number of surf trails start from search engines and
are very short, nonnegligible surfing may still be occurring without
support from search engines. We use the following model for mixed surfing and searching:

\begin{itemize}
\item While performing {\em random surfing}~\cite{pagerank}, users traverse a link to some
neighbor with probability $(1-c)$, and jump to a random page with
probability $c$.  The constant $c$ is known as the {\em teleportation
probability}, typically set to 0.15~\cite{glen}.

\item While browsing the Web, users perform random surfing with
probability $x$. With probability $(1-x)$ users query a search
engine and browse among results presented in the form of a ranked list.
\end{itemize}

We still assume that there is only one search engine that every user
uses for querying. However, this assumption does not significantly restrict 
the applicability of our model. 
For our purposes the effect of multiple search engines that present
the same ranked list for a query is equivalent to a single search
engine that presents the same ranked list and gets a user traffic
equal to the sum of the user traffic of the multiple search engines.

Assuming that page popularity is measured using PageRank, 
under our mixed browsing model the expected visit rate of a
page $p$ at time $t$ is given by:
\begin{eqnarray*}
V(p,t) &=& (1-x) \cdot F(P(p,t)) \\
&+& x \cdot \bigg(\big((1-c) \cdot \frac{P(p,t)}{\sum_{p' \in 
\mathcal{P}}P(p',t)} + c \cdot \frac{1}{n} \big) \bigg)
\end{eqnarray*}

\begin{figure}[!t]
\begin{center}
\includegraphics[width=200pt]{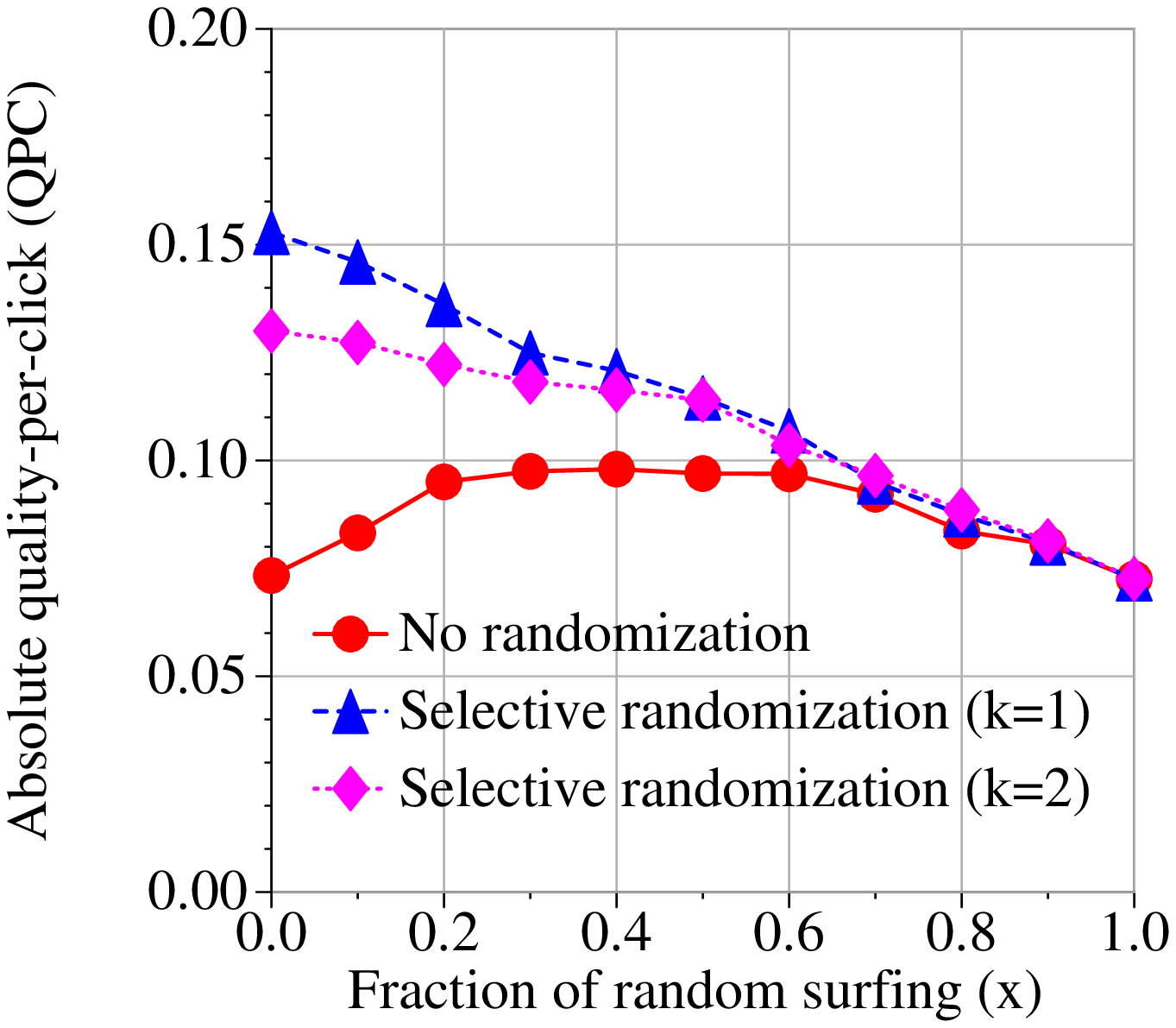}
\caption{Influence of the extent of random surfing.}
\label{fig:mixed}
\end{center}
\end{figure}

Figure~\ref{fig:mixed} shows 
absolute QPC values for different values of $x$ (based on simulation). 
Unlike with other
graphs in this paper, in this graph we plot the absolute value of QPC,
because the ideal QPC value varies with the extent of random
surfing ($x$). Recall that $x=0$ denotes pure search engine
based surfing, while $x=1$ denotes pure random surfing. Observe that for all values of 
$x$, randomized rank promotion performs better than (or as well as) nonrandomized
ranking. It is interesting to observe that when $x$ is small, 
random surfing helps nonrandomized ranking, since random surfing
increases the chances of exploring unpopular pages (due to the teleportation
probability). However, beyond a certain extent, it does not help as much as it 
hurts (due to the exploration/exploitation tradeoff as was the case for randomized rank promotion).


\section{Summary}
\label{sec:concl}

The standard method of ranking search results deterministically according to popularity
has a significant flaw: high-quality Web pages that happen to be new are drastically undervalued. 
In this paper we first presented results of a real-world study 
which demonstrated that diminishing the bias against new pages by
selectively and transiently promoting them in rank can improve overall
result quality substantially.
We then showed through extensive simulation 
of a wide variety of Web community types
that promoting new pages by partially randomizing
rank positions (using just $10\%$ randomization) consistently
leads to much higher-quality search results
compared with strict deterministic ranking. 
From our empirical results we conclude that randomized rank promotion
is a promising approach that merits further study and evaluation.
To pave the way for further work, we have developed new analytical
models of Web page popularity evolution under
deterministic and randomized search result ranking,
and introduced formal metrics by which to evaluate ranking methods.

\begingroup \raggedright \small
\bibliographystyle{abbrv}

\endgroup

\vspace{-6pt}

\appendix

\section{Real-World Effectiveness of Rank Promotion}
\label{sec:jokes}

In this section we describe a live experiment we conducted to study the 
effect of rank promotion on the
evolution of popularity of Web pages.

\subsection{Experimental Procedure}
For this experiment we created our own small Web community consisting of several thousand
Web pages containing entertainment-oriented content, and nearly one thousand volunteer users who had 
no prior knowledge of this project.

\vspace{9pt}
\noindent
{\bf Pages:} We focused on entertainment because we felt it would be relatively easy to attract a large number of users.
The material we started with consisted of a large number of jokes gathered from online databases.
We decided to use ``funniness'' as a surrogate
for quality, since users are generally willing to provide their opinion about how funny something is.
We wanted the funniness distribution of our jokes to mimic the quality distribution of pages on the Web. 
As far as we know PageRank is the best available estimate of the quality distribution of Web pages,
so we downsampled our initial collection of jokes and quotations
to match the PageRank distribution reported in~\cite{impact}. To determine the funniness of our jokes for this purpose 
we used numerical user ratings provided by the source databases. Since most Web pages
have very low PageRank, we needed a large number of nonfunny items to match 
the distribution, so we chose to supplement jokes with quotations. We obtained our 
quotations from sites offering insightful quotations not intended to be humorous. 
Each joke and quotation was converted into a single Web page on our site.

\vspace{9pt}
\noindent
{\bf Overall site:} The main page of the Web site we set up consisted of 
an ordered list of links to individual joke/quotation
pages, in groups of ten at a time, as is typical in search engine responses. Text at the top stated that the 
jokes and quotations were presented
in descending order of funniness, as rated by users of the site. Users 
had the option to rate the items: we equipped each joke/quotation page 
with three buttons, labeled ``funny,'' ``neutral,'' and ``not funny.'' To 
minimize the possibility of voter fraud, 
once a user had rated an item the buttons were removed from that item, and 
remained absent upon all subsequent visits by the same user to the same page.

\vspace{9pt}
\noindent
{\bf Users:} We advertised our site daily over a period of $45$ days, and encouraged visitors to rate whichever jokes and quotations they decided to view. Overall we had $962$ participants.
Each person who visited the site for the first time was assigned at random
into one of two user groups (we used cookies to ensure consistent group membership across multiple visits,
assuming few people would visit our site from multiple computers): 
one group for which rank promotion was used,
and one for which rank promotion was not used. 
For the latter group, items were presented in descending order of current popularity,
measured as the number of funny votes submitted by members of the 
group.\footnote{Due to the relatively small scale of our experiment 
there were frequent ties in popularity values. We chose to break ties 
based on age, with older pages receiving better rank positions, to 
simulate a less discretized situation.}
For the other group of users, items were also presented in descending order of popularity among
members of the group,
except that all items that had not yet been viewed by any user were inserted 
in a random order starting at rank position $21$ (This variant corresponds to selective 
promotion with $k=21$ and $r=1$.). 
A new random order for these zero-awareness items was chosen for each unique 
user. Users were not informed that rank promotion was
being employed.

\vspace{9pt}
\noindent
{\bf Content rotation:} For each user group we kept the number of accessible 
joke/quotation items fixed at $1000$ throughout the duration of our $45$-day 
experiment. However, each item had a finite lifetime of less than $45$ days. 
Lifetimes for the initial $1000$ items were assigned uniformly at random from 
$[1,30]$, to simulation a steady-state situation
in which each item had a real lifetime of $30$ days.
When a particular item expired we replaced it with another item of the same 
quality, and set its lifetime to $30$ days and its initial popularity to zero. 
At all times we used the same joke/quotation items for both user groups.

\subsection{Results}
First, to verify that the subjects of our experiment behaved similarly to users of a search engine, 
we measured the relationship between the rank of an item and the number of user visits it received. We discovered
a power-law with an exponent remarkably close to $-3/2$, which is precisely 
the relationship between rank and number of visits
that has been measured from usage logs of the AltaVista search engine (see Section~\ref{sec:g} for details).

We then proceeded to assess the impact of rank promotion. 
For this purpose we wanted to analyze a steady-state scenario, so we only measured the outcome of the
final $15$ days of our experiment (by then all the original items had expired and been replaced).
For each user group we measured the ratio of funny votes to total votes during this period.
Figure~\ref{fig:intbarQPC} shows the result.  The ratio achieved using rank promotion
was approximately $60\%$ larger than that obtained using
strict ranking by popularity.

\section{Proof of Theorem 1}
\label{apx:thm1Proof}

  Because we consider only the pages of quality $q$
  and we focus on steady-state behavior,
  we will drop $q$ and $t$ from our notation unless it causes confusion.
  For example, we use $f(a)$ and $V(p)$
  instead of $f(a|q)$ and $V(p,t)$ in our proof.

  We consider a very short time interval $dt$ during which
  every page is visited by at most one monitored user. That is, 
  $V(p) dt < 1$ for every page $p$. Under this assumption
  we can interpret $V(p) dt$ as the probability that
  the page $p$ is visited by one monitored user during the time interval $dt$.

  Now consider the pages of awareness $a_i = \frac{i}{m}$. 
  Since these pages are visited by at most one monitored user during $dt$, their
  awareness will either stay at $a_i$ or increase to $a_{i+1}$.
  We use $\mathcal{P}_S(a_i)$ and $\mathcal{P}_I(a_i)$ to denote the probability that 
  that their awareness remains at $a_i$ or
  increases from $a_i$ to $a_{i+1}$, respectively.
  The awareness of a page increases
  if a monitored user who was previously unaware of the page visits it. 
  The probability that a monitored user visits $p$
  is $V(p) dt$. The probability that a random monitored user
  is aware of $p$ is $(1-a_i)$. Therefore,

  \begin{align}
    \label{eq:PI}
    \mathcal{P}_I(a_i) &= V(p) dt (1-a_i) = F(P(p)) dt (1-a_i)\notag\\
             &= F(q a_i) dt (1-a_i)
  \end{align}
  Similarly, 
  \begin{equation}
    \label{eq:PS}
    \mathcal{P}_S(a_i) = 1 - \mathcal{P}_I(a_i) = 1 - F(q a_i) dt (1-a_i)
  \end{equation}

  We now compute the fraction of pages
  whose awareness is $a_i$ after $dt$. 
  We assume that before $dt$, $f(a_i)$ and $f(a_{i-1})$ 
  fraction of pages have awareness $a_i$ and $a_{i-1}$, respectively.
  A page will have awareness
  $a_i$ after $dt$ if (1) its awareness is $a_i$ before $dt$ and 
  the awareness stays the same
  or (2) its awareness is $a_{i-1}$ before $dt$, but it increases to $a_{i}$.
  Therefore, the fraction of pages at awareness $a_i$ after $dt$
  is potentially
  \begin{equation*}
    f(a_i)\mathcal{P}_S(a_{i}) + f(a_{i-1})\mathcal{P}_I(a_{i-1}).
  \end{equation*}
  However, under our Poisson model, a page disappears 
  with probability $\lambda dt$ during the time interval $dt$.
  Therefore, only $(1 - \lambda dt)$ fraction will survive 
  and have awareness $a_i$ after $dt$:
  \begin{equation*}
    [f(a_i)\mathcal{P}_S(a_{i}) + f(a_{i-1})\mathcal{P}_I(a_{i-1})](1-\lambda dt)
  \end{equation*}
  Given our steady-state assumption, the fraction of pages
  at $a_i$ after $dt$ is the same as the fraction of pages
  at $a_i$ before $dt$. Therefore,
  \begin{equation}\label{eq:fa}
    f(a_i) = [f(a_i)\mathcal{P}_S(a_{i}) + f(a_{i-1})\mathcal{P}_I(a_{i-1})](1-\lambda dt).
  \end{equation}
  From Equations~\ref{eq:PI}, \ref{eq:PS} and~\ref{eq:fa}, we get
  \begin{equation}
    \frac{f(a_i)}{f(a_{i-1})} 
      = \frac{(1-\lambda dt) F(q a_{i-1}) dt (1-a_{i-1})}{
        (\lambda + F(q a_i)) dt (1-a_i)}\notag
  \end{equation}
  Since we assume $dt$ is very small, we can ignore the second
  order terms of $dt$ in the above equation and simplify it to
  \begin{equation}
    \frac{f(a_i)}{f(a_{i-1})} 
      = \frac{F(q a_{i-1}) (1-a_{i-1})}{
        (\lambda + F(q a_i)) (1-a_i)}
  \end{equation}
  From the multiplication of 
  $\frac{f(a_i)}{f(a_{i-1})} \times
  \frac{f(a_{i-1})}{f(a_{i-2})} \times \dots \times
  \frac{f(a_{1})}{f(a_{0})}$,
  we get
  \begin{equation}\label{eq:recurrence}
    \frac{f(a_i)}{f(a_{0})} 
      = \frac{1-a_0}{1-a_i} 
      \prod_{j=1}^{i} \frac{F(q a_{j-1})}{\lambda + F(q a_j)}
  \end{equation}
  
  We now compute $f(a_0)$. Among the pages
  with awareness $a_0$, $\mathcal{P}_S(a_0)$ fraction will stay at $a_0$ after $dt$.
  Also, $\lambda dt$ fraction new pages will appear, and their awareness is $a_0$
  (recall our assumption that new pages start with zero awareness).
  Therefore,
  \begin{equation}
    f(a_0) = f(a_0)\mathcal{P}_S(a_0)(1-\lambda dt) + \lambda dt
  \end{equation}
  After rearrangement and ignoring the second order terms of $dt$, we get
  \begin{equation}
    \label{eq:a0}
    f(a_0) = \frac{\lambda}{F(q a_0) + \lambda} 
    = \frac{\lambda}{F(0) + \lambda}
  \end{equation}
  By combining Equations~\ref{eq:recurrence} and~\ref{eq:a0}, we get
  \begin{align*}
    f(a_i)
      &= f(a_0)\frac{1-a_0}{1-a_i} 
      \prod_{j=1}^{i} \frac{F(q a_{j-1})}{\lambda + F(q a_j)}\\
      & = \frac{\lambda}{(\lambda + F(0)) (1-a_{i})} \prod_{j=1}^{i} 
    \frac{F(q a_{j-1})}{\lambda + F(q a_{j})}
  \end{align*}

\end{document}